\documentclass[preprint]{ptephy_v1}%%%%%% to generate preprint number with ptep logo

\preprintnumber{XXXX-XXXX} %%% %%% Insert preprint number here
\usepackage{amsmath,amssymb,amsthm}
\usepackage{booktabs}
\usepackage{graphicx}
\usepackage{hyperref}
\newcommand{\Sn}{S_n}

\newcommand{\id}{\mathrm{id}}
\newcommand{\Irr}{\operatorname{Irr}}
\newcommand{\e}{\mathrm{e}}

\newcommand{\cont}{\operatorname{cont}}

\theoremstyle{definition}

\begin{document}

\title{Cavity method for permutation models on Cayley trees}

\author{Masayuki Ohzeki}
\affil{Department of Physics, Institute of Science Tokyo, Tokyo, 152-8551, Japan  \\
Graduate School of Information Sciences, Tohoku University, Miyagi 980-8564, Japan \\
Research and Education Institute for Semiconductors and Informatics, Kumamoto University, Kumamoto 860-8555, Japan \\
Sigma-i Co., Ltd., Tokyo, 108-0075, Japan
\email{mohzeki@tohoku.ac.jp}}

\begin{abstract}
Motivated by permutation statistical models arising in random tensor networks, we study permutation models on a Cayley tree whose variables take values in the symmetric group $\Sn$.
The pair interaction is assumed to depend only on the cycle type of the relative permutation.
Then the Boltzmann weight is written as a class function on $\Sn$.
This property diagonalizes the edge convolution operator in irreducible representation sectors.
As a result, the linear stability of the uniform paramagnetic cavity solution is controlled by the character eigenvalue ratios.
For cycle-factorized weights, these eigenvalues can be expressed as specializations of Schur functions.
We derive the instability criteria and also verify their validity by comparison with direct numerical iterations of the cavity equation.
\end{abstract}

\subjectindex{A13, A40, A42}

\maketitle

\section{Introduction}

Statistical models with degrees of freedom in the symmetric group have recently become important in quantum information science and high-energy theory.
In random quantum circuits and random tensor networks (RTNs), the replica method maps moments of entanglement entropy to classical statistical models whose local variables are permutations~\cite{Zhou2019,Romain2019,Yimu2020,Jian2020,Hayden2016}.
These mappings bring permutation models, which were already studied in earlier statistical-mechanical contexts~\cite{Drouffe1979}, into the analysis of entanglement transitions and holographic toy models.

A central problem in such models is the determination of critical points.
Duality analysis provides one of the most useful guiding principles in finite-dimensional lattice models.
The Kramers--Wannier duality of the two-dimensional Ising model locates the exact critical point without explicitly calculating the free energy~\cite{Kramers1941}.
Fourier transformation extends the duality to $Z_n$ models~\cite{WuWang1976}, and even for spin glasses treated by the replica method~\cite{Nishimori1979,Nishimori2002,Maillard2003,Nishimori2006,Ohzeki2008hl,Ohzeki2009,Ohzeki2011slope,Ohzeki2015}, quantum error-correcting codes~\cite{Dennis2002,Ohzeki2012duality,Ohzeki2012,Bombin2014}, and ground-state problems~\cite{Ohzeki2013,Ohzeki2018gs,Miyazaki2020}.
Even when self-duality is absent, a single equation often gives reasonable conjectures for critical points, especially when combined with real-space renormalization analysis~\cite{Ohzeki2009,Ohzeki2015}.
For the permutation model of RTNs, a recent application of such a single equation yielded an explicit estimate of the critical bond dimension separating distinct entanglement regimes~\cite{Ohzeki2024}.
However, such criteria do not, by themselves, provide a systematic analytic solution to the finite-dimensional permutation model.

This situation motivates the development of complementary calculational tools based on standard methods in statistical mechanics.
Mean-field theory, the Bethe approximation, cavity methods, and Migdal--Kadanoff-type approximations have long been used as quantitative tools for ordinary spin models~\cite{Bethe1935,Kadanoff1976,Nishimori2010}.
For permutation models relevant to RTNs, however, explicit Bethe or Migdal--Kadanoff calculations are still limited.
The main obstacle is algebraic: the spin variables take values in a non-Abelian group, and the natural transfer kernels live in the group algebra of $\Sn$.
Therefore, any systematic calculation must use noncommutative Fourier analysis, character theory, and symmetric functions.

There is also a closely related mean-field approach to RTN entanglement transitions based on random tree tensor networks~\cite{LopezPiqueres2020}.
In that work, the entanglement problem was mapped to a replica statistical-mechanical model on a Cayley tree and analyzed by the cavity method, giving a tractable mean-field example of an entanglement transition between different entanglement-scaling regimes.
The present work is complementary.
Rather than starting from the entanglement of a tree tensor network itself, we focus on the homogeneous cavity recursion of the finite-$n$ symmetric-group permutation spin model and diagonalize its bulk instability by representation theory.
The purpose of this paper is to analyze the same class of permutation interactions from a complementary point of view in which the cavity equation is exact.
On a Cayley tree, the transition from the uniform paramagnetic message can be studied directly by linearizing the cavity recursion.
Because the RTN edge weight and its natural generalizations are class functions on the symmetric group, the convolution part of the cavity equation is diagonalized by the noncommutative Fourier transform.
The relevant eigenmodes are labeled by Young diagrams, and the instability condition reduces to a comparison of character eigenvalues.

In this paper, we formulate the cavity method for permutation models on a Cayley tree and compute the linear instability of the paramagnetic solution in representation space.
The calculation proceeds as follows.
Section~2 introduces the permutation model motivated by RTNs and fixes the notation for the cycle-count interaction.
Section~3 derives the homogeneous cavity equation on the Cayley tree and diagonalizes the edge convolution by using the noncommutative Fourier transform and the character theory of $\Sn$.
Section~4 applies the resulting stability criterion to several examples, including the Ising limit, the complete-matching interaction, fixed-point-enhanced interactions, and the RTN cycle-count model.
Section~5 verifies the analytic transition points by direct numerical iteration of the cavity equation.
The last section is devoted to the conclusion.

\section{Problem setting}
The permutation model is closely related to RTNs, which are random quantum states defined on networks.
After averaging over random tensors, the replica method maps the calculation of entanglement entropies to differences of free energies of permutation models with appropriate boundary conditions~\cite{Hayden2016,Romain2019}.
In the present notation, this order is denoted by $n$.
The physical RTN limit is obtained by analytically continuing the final expressions toward $n\to0$.
In the calculations below we first keep $n$ as a positive integer so that the representation theory of $\Sn$ is well defined, and only afterwards discuss the implication of the replica limit.
The transition between volume-law and area-law scaling of entanglement entropy in RTNs corresponds, in this effective classical model, to the transition between a ferromagnetic phase and a paramagnetic phase of the permutation spins.
On a graph $G=(V,E)$, the cycle-count Hamiltonian is
\begin{equation}
  H(\boldsymbol{\sigma})
  =
  -\sum_{(i,j)\in E} C(\sigma_i\sigma_j^{-1}),
  \label{eq:hamiltonian_permutation_spin}
\end{equation}
where a spin variable $\sigma_v$ at each vertex $v\in V$ is an element of the symmetric group and $C$ is the cycle-counting function.
For example, the identity permutation has $C(e)=n$, while the permutation $(1,2)(3)$ in $S_3$ has two cycles.
The cycle-counting function satisfies
\begin{equation}
  C(\sigma^{-1})=C(\sigma),
  \label{eq:cycle_inverse}
\end{equation}
and it is invariant under conjugation $C(\tau^{-1}\sigma\tau)=C(\sigma)$.
The interaction energy is invariant under global left or right multiplication of all spins by a common group element.
This symmetry and the class-function property are the basic algebraic inputs used below.

The basic RTN edge weight is
\begin{equation}
  f_D(g)=D^{C(g)},
  \label{eq:rtn_weight}
\end{equation}
where $C(g)$ denotes the total number of cycles in $g\in\Sn$.
On a graph $G=(V,E)$ the corresponding partition function is
\begin{equation}
  Z_G(D)
  =
  \sum_{\{\pi_i\in\Sn\}}
  \prod_{\langle ij\rangle\in E}
  f_D(\pi_i\pi_j^{-1}).
  \label{eq:rtn_partition}
\end{equation}
More generally, the analysis below applies to any class-function interaction
\begin{equation}
  W(g)=\exp\{\beta J(g)\}.
\end{equation}
We also assume conjugation invariance, $W(h^{-1}gh)=W(g)$.
The RTN weight in Eq.~\eqref{eq:rtn_weight} is recovered by setting $\beta =\log D$ for the cycle-count energy $J(g)=C(g)$.

\section{Method}
\subsection{Cavity method}
Here we formulate the cavity recursion on the Cayley tree.
We define the number of incoming descendant messages used to form a cavity message sent toward the root as $z-1$, where the vertex degree is denoted by $z$.
At each vertex $i$ we place a spin $\pi_i\in\Sn$.

Let the cavity message along the directed edge $i\to j$ be
\begin{equation}
  \eta_{i\to j}\colon \Sn\to\mathbb{R}_{\ge 0},
  \qquad
  \sum_{\pi\in\Sn}\eta_{i\to j}(\pi)=1
\end{equation}
Since the underlying graph is a tree, the exact message update is
\begin{equation}
  \eta_{i\to j}(\pi)
  =
  \frac{1}{Z_{i\to j}}
  \prod_{\ell\in\partial i\setminus j}
  \left[
    \sum_{\sigma\in\Sn}
    W(\pi^{-1}\sigma)\eta_{\ell\to i}(\sigma)
  \right].
  \label{eq:directed_cavity}
\end{equation}
On a homogeneous tree, if all incoming messages in a generation are identical, the recursion reduces to a one-dimensional iteration on functions on $\Sn$.
Define the edge convolution operator
\begin{equation}
  (T_W f)(\pi)
  =
  \sum_{\sigma\in\Sn}W(\pi^{-1}\sigma)f(\sigma).
\end{equation}
Then the homogeneous cavity equation is
\begin{equation}
  \eta_{t+1}(\pi)
  =
  \frac{(T_W\eta_t(\pi))^{z-1}}
       {\sum_{\tau\in\Sn}(T_W\eta_t(\tau))^{z-1}}.
  \label{eq:cavity}
\end{equation}
We analyze the linear stability of the uniform solution of this map.
The rest of the calculation starts from Eq.~\eqref{eq:cavity}: the edge convolution is first diagonalized in representation space, and the nonlinear cavity map is then linearized around the uniform solution.

\subsection{Noncommutative Fourier transform}

This section performs the algebraic diagonalization of the edge convolution operator $T_W$.
For an Abelian spin space this step would be an ordinary Fourier transform.
For the non-Abelian group $\Sn$, the Fourier components are labeled by irreducible representations, and class functions act as scalars in each irreducible sector.

Let $\lambda\vdash n$ label an irreducible representation of $\Sn$.
We denote the representation matrix by $\rho^\lambda$, its dimension by $d_\lambda$, and its character by $\chi^\lambda$.
For a function $f$ on $\Sn$, we use the finite-group noncommutative Fourier transform
\begin{equation}
  \widehat f(\lambda)
  =
  \sum_{g\in\Sn}f(g)\rho^\lambda(g).
\end{equation}
With this convention, the inverse transform is
\begin{equation}
  f(g)
  =
  \frac{1}{n!}
  \sum_{\lambda\vdash n}
  d_\lambda
  \operatorname{Tr}
  \left[
    \widehat f(\lambda)\rho^\lambda(g^{-1})
  \right].
\end{equation}
For a class function $W$, define
\begin{equation}
  A_\lambda
  =
  \sum_{g\in\Sn}W(g)\rho^\lambda(g).
\end{equation}
Since $W$ is a class function, $A_\lambda$ commutes with every $\rho^\lambda(h)$.
By Schur's lemma, $A_\lambda$ becomes a scalar matrix:
\begin{equation}
  A_\lambda=\omega_\lambda I_{d_\lambda}.
\end{equation}
Taking the trace gives
\begin{equation}
  \omega_\lambda
  =
  \frac{1}{d_\lambda}
  \sum_{g\in\Sn}W(g)\chi^\lambda(g)
  ,
  \label{eq:eigenvalue_general}
\end{equation}
where $\chi^\lambda(g)$ is the character of the irreducible representation $\lambda$ evaluated at $g$.
Thus the edge convolution is diagonalized by irreducible representations.
The tools used here are the finite-group Fourier transform, Schur's lemma, and the character formula for a central group-algebra element.
They are summarized in Appendix~\ref{app:finite_group_fourier}.

We recall the conjugacy classes of $\Sn$, which are labeled by partitions
\begin{equation}
  \mu=(1^{m_1}2^{m_2}\cdots)\vdash n .
\end{equation}
Let $C_\mu$ be the corresponding conjugacy class.
The standard formula for its size is
\begin{equation}
    |C_\mu|=\frac{n!}{z_\mu},
\end{equation}
where
\begin{equation}
  z_\mu=\prod_{k\ge 1}k^{m_k}m_k!.
\end{equation}
This formula follows from the centralizer size of a permutation of cycle type $\mu$; see Appendix~\ref{app:symmetric_group_classes}.
If $W_\mu$ denotes the value of $W$ on $C_\mu$, Eq.~\eqref{eq:eigenvalue_general} is rewritten as
\begin{equation}
  \omega_\lambda
  =
  \frac{1}{d_\lambda}
  \sum_{\mu\vdash n}
  \frac{n!}{z_\mu}
  W_\mu\chi^\lambda_\mu
  .
  \label{eq:eigenvalue_class}
\end{equation}
For the trivial representation $[n]$, $\chi^{[n]}_\mu=1$, hence
\begin{equation}
  \omega_{[n]}
  =
  \sum_{\mu\vdash n}\frac{n!}{z_\mu}W_\mu
  =
  \sum_{g\in\Sn}W(g)
  .
  \label{eq:omega_trivial}
\end{equation}
For an arbitrary cycle function $F$, written as
\begin{equation}
  J(g)=F(\mu(g)),
  \qquad
  W(g)=\e^{\beta F(\mu(g))},
\end{equation}
we obtain
\begin{equation}
  \omega_\lambda(\beta)
  =
  \frac{1}{d_\lambda}
  \sum_{\mu\vdash n}
  \frac{n!}{z_\mu}
  \e^{\beta F(\mu)}
  \chi^\lambda_\mu
  .
  \label{eq:arbitrary_cycle_function}
\end{equation}
This is the basic formula for general cycle-type interactions.
The purpose of the following sections is to evaluate this eigenvalue formula in concrete models and substitute the resulting ratios into the cavity stability condition.

\subsection{Linear stability analysis}
This section turns the diagonalization of the edge convolution into a transition criterion.
We perturb the uniform cavity message, project the perturbation onto an irreducible representation sector, and compute whether that perturbation grows or decays after one cavity update.
The uniform distribution
\begin{equation}
  \eta_0(\pi)=\frac{1}{n!}
\end{equation}
is trivially a fixed point of Eq.~\eqref{eq:cavity}, which corresponds to the paramagnetic phase.
Write a small perturbation as
\begin{equation}
  \eta_t(\pi)=\frac{1}{n!}+\delta_t(\pi),
\end{equation}
where $\sum_{\pi\in\Sn}\delta_t(\pi)=0$.
The normalization condition eliminates the trivial-representation component.
If $\delta_t$ belongs to a nontrivial irreducible sector $\lambda\ne[n]$, linearization of Eq.~\eqref{eq:cavity} yields
\begin{equation}
  \delta_{t+1}
  =
  (z-1)\frac{\omega_\lambda}{\omega_{[n]}}\delta_t
  +O(\delta_t^2).
  \label{eq:linearized}
\end{equation}
Therefore the local stability condition for the uniform solution is
\begin{equation}
  (z-1)
  \max_{\lambda\ne[n]}
  \left|
  \frac{\omega_\lambda}{\omega_{[n]}}
  \right|
  <1.
  \label{eq:stability}
\end{equation}
When a ferromagnetic bifurcation is dominated by a sector $\lambda$, its instability condition is
\begin{equation}
  (z-1)\frac{\omega_\lambda}{\omega_{[n]}}=1.
  \label{eq:critical_mode}
\end{equation}
In the examples below, the standard representation $[n-1,1]$ is the natural order-parameter sector.
The derivation of Eq.~\eqref{eq:linearized} is detailed in Appendix~\ref{app:cavity_linearization}.
Equation~\eqref{eq:stability} gives the local stability of the uniform solution.
For a continuous transition, this condition coincides with the thermodynamic transition point.
For Potts-type models, however, a first-order transition may occur before the spinodal instability of the uniform solution; in that case one must compare Bethe free energies separately.
In the present study, we focus on cases in which the transition is detected by this continuous instability of the paramagnetic solution.

\paragraph{$S_2$ and the Ising model}
We check the validity of our formalism for several simple examples.
The two representation sectors of $S_2$ reproduce the symmetric and antisymmetric eigenmodes of the usual Ising transfer matrix.

Let $S_2=\{e,\tau\}$, where $e$ is the identity.
The Ising model can be represented by
\begin{equation}
  W(e)=\e^{\beta},
  \qquad
  W(\tau)=\e^{-\beta}.
\end{equation}
The irreducible representations of $S_2$ are the trivial representation $[2]$ and the sign representation $[1,1]$.
The corresponding eigenvalues are
\begin{equation}
  \omega_{[2]}
  =
  \e^{\beta}+\e^{-\beta}
  =
  2\cosh(\beta),
\end{equation}
\begin{equation}
  \omega_{[1,1]}
  =
  \e^{\beta}-\e^{-\beta}
  =
  2\sinh(\beta).
\end{equation}
Hence
\begin{equation}
  \frac{\omega_{[1,1]}}{\omega_{[2]}}
  =
  \tanh(\beta),
\end{equation}
and the critical condition is
\begin{equation}
  (z-1)\tanh(\beta_c)=1.
\end{equation}
This is well known as the critical point on the Cayley tree of the standard Ising model.

\paragraph{Complete-matching interaction}
Next, we take the Potts-like interaction that rewards exact equality of neighboring permutations.
It is a useful benchmark because all nontrivial irreducible sectors become degenerate.

Consider the interaction that gives an enhanced weight only when neighboring permutations match exactly:
\begin{equation}
  W(g)=\exp(\beta\delta_{g,e}).
\end{equation}
Let $r=\e^{\beta}$.
Then
\begin{equation}
  W(e)=r,
  \qquad
  W(g)=1\quad(g\ne e).
\end{equation}
For the trivial representation,
\begin{equation}
  \omega_{[n]}=r+n!-1.
\end{equation}
For a nontrivial irreducible representation $\lambda\ne[n]$, character orthogonality gives
\begin{equation}
  \sum_{g\in\Sn}\chi^\lambda(g)=0,
  \qquad
  \chi^\lambda(\id)=d_\lambda.
\end{equation}
Thus
\begin{equation}
  \omega_\lambda
  =
  \frac{1}{d_\lambda}
  \left(rd_\lambda+\sum_{g\ne\id}\chi^\lambda(g)\right)
  =
  r-1.
\end{equation}
All nontrivial sectors are degenerate, and
\begin{equation}
  \frac{\omega_\lambda}{\omega_{[n]}}
  =
  \frac{r-1}{r+n!-1}
  \qquad(\lambda\ne[n]).
\end{equation}
The instability condition is
\begin{equation}
  (z-1)\frac{r_c-1}{r_c+n!-1}=1,
\end{equation}
or
\begin{equation}
  \e^{\beta_c}
  =
  \frac{(z-1)+n!-1}{(z-1)-1}.
  \label{eq:delta_critical}
\end{equation}
This is identical to the spinodal condition of the $n!$-state Potts model on the Cayley tree.

\paragraph{Cycle-factorized interactions}
We take a general class of cycle-factorized weights.
The main point is that the character sums can be evaluated by Frobenius' characteristic map and expressed as Schur-function specializations.

Let $c_k(g)$ denote the number of cycles of length $k$ in $g\in\Sn$.
We define
\begin{equation}
  J(g)=\sum_{k=1}^{n}\epsilon_k c_k(g).
\end{equation}
Then the Boltzmann weight factorizes as
\begin{equation}
  W(g)=\prod_{k=1}^{n}x_k^{c_k(g)},
\end{equation}
where $x_k=\e^{\beta\epsilon_k}$.
It is straightforward to obtain the following expression.
\begin{equation}
  \omega_{\lambda} = \frac{1}{d_{\lambda}}
  \sum_{g\in\Sn}
  \chi^\lambda(g)
  \prod_{k=1}^{n}x_k^{c_k(g)}.
\end{equation}
By Frobenius' characteristic formula, this is equivalently written as
\begin{equation}
  \omega_\lambda
  =
  \frac{n!}{d_\lambda}s_\lambda[x_k].
  \label{eq:schur_eigenvalue}
\end{equation}
\paragraph{Fixed-point-enhanced model}
We apply the Schur-function formula to a model that favors fixed points of the relative permutation.
The standard representation is the natural candidate for the leading ferromagnetic instability because its character is $c_1(g)-1$.

Consider the model
\begin{equation}
  W(g)=x^{c_1(g)},
\end{equation}
where $x=\e^{\beta}$.
This corresponds to the specialization $\epsilon_1=1$, and $\epsilon_k=0\quad(k\ge 2)$.
For the trivial representation, $s_{[n]}=h_n$.
Using the generating function of complete homogeneous symmetric functions,
\begin{equation}
  \sum_{m=0}^{\infty}h_m z^m
  =
  \exp\left(\sum_{k=1}^{\infty}\frac{x_k}{k}z^k\right)
  =
  \frac{\e^{(x-1)z}}{1-z},
\end{equation}
we obtain
\begin{equation}
  h_m(x)=\sum_{a=0}^{m}\frac{(x-1)^a}{a!}.
\end{equation}
Thus
\begin{equation}
  \omega_{[n]}=n!h_n(x).
\end{equation}

For the standard representation $[n-1,1]$, we find $d_{[n-1,1]}=n-1$.
By the Jacobi--Trudi formula,
\begin{equation}
  s_{[n-1,1]}(x)=h_{n-1}(x)h_1(x)-h_n(x)=xh_{n-1}(x)-h_n(x).
\end{equation}
Hence
\begin{equation}
  \omega_{[n-1,1]}
  =
  \frac{n!}{n-1}
  \left(xh_{n-1}(x)-h_n(x)\right).
\end{equation}
The critical condition is thus 
\begin{equation}
  (z-1)
  \frac{(x_c-1) h_{n-1}(x_c)-(x_c-1)^n/n!}{(n-1)h_n(x_c)}
  =
  1.
  \label{eq:fixed_point_critical}
\end{equation}
The generating-function computation and the Jacobi--Trudi step are detailed in Appendix~\ref{app:fixed_point_calc}.

For example, for $n=3$ and $z=3$, the critical equation is
\begin{equation}
  (x_c-1)^3+3(x_c-1)^2-6=0,
\end{equation}
whose positive solution is $x_c\simeq 2.195823$, and $\beta_c\simeq 0.7865570766$.

\paragraph{Cycle-count model}
We take the RTN cycle-count weight.
Because all cycle lengths carry the same fugacity $D$, the hook-content formula gives a closed product expression for each representation-sector eigenvalue.

Let
\begin{equation}
  c(g)=\sum_{k=1}^{n}c_k(g)
\end{equation}
be the total number of cycles in $g$.
We consider
\begin{equation}
  W(g)=D^{c(g)},
  \qquad
  D=\e^{\beta}.
\end{equation}
Then $p_k=D$ for all $k$.
The Schur specialization is evaluated by the hook-content formula:
\begin{equation}
  s_\lambda[p_k=D]
  =
  \frac{\prod_{u\in\lambda}(D+\cont(u))}
       {\prod_{u\in\lambda}h(u)}.
\end{equation}
Here $u=(i,j)$ is a box of the Young diagram, $\cont(u)=j-i$, and $h(u)$ is the hook length.
The hook-length formula gives
\begin{equation}
  d_\lambda
  =
  \frac{n!}{\prod_{u\in\lambda}h(u)}.
\end{equation}
Combining this with Eq.~\eqref{eq:schur_eigenvalue}, we get
\begin{equation}
  \omega_\lambda
  =
  \prod_{u\in\lambda}\bigl(D+\cont(u)\bigr).
  \label{eq:hook_content_eigenvalue}
\end{equation}
For the trivial representation $[n]$, the contents are $0,1,\ldots,n-1$, hence
\begin{equation}
  \omega_{[n]}
  =
  D(D+1)\cdots(D+n-1).
\end{equation}
On the other hand, for the standard representation $[n-1,1]$, the contents are
\begin{equation}
  0,1,\ldots,n-2,-1.
\end{equation}
Thus
\begin{equation}
  \omega_{[n-1,1]}
  =
  (D-1)D(D+1)\cdots(D+n-2).
\end{equation}
Then the critical condition becomes
\begin{equation}
  (z-1)\frac{D_c-1}{D_c+n-1}=1,
\end{equation}
or equivalently
\begin{equation}
  D_c=\frac{z+n-2}{z-2}.
  \label{eq:cycle_count_critical}
\end{equation}
The combination of the hook-content formula and the hook-length formula is detailed in Appendix~\ref{app:cycle_count_calc}.

\subsection{Bulk magnetization}
The same formula also gives a simple Bethe approximation estimate for regular finite-dimensional lattices.
If the coordination number of the lattice is $z$, the Bethe approximation replaces the local environment by a tree with branching number $z-1$.
For the RTN cycle-count model, Eq.~\eqref{eq:cycle_count_critical} therefore gives, for the square lattice, $z=4$,
\begin{equation}
  D_c^{\mathrm{Bethe}}(z=4,n)
  =
  \frac{n+2}{2}.
  \label{eq:square_lattice_bethe_estimate}
\end{equation}
The RTN problem requires the replica limit $n\to0$.
Taking this limit in Eq.~\eqref{eq:square_lattice_bethe_estimate} gives
\begin{equation}
  \lim_{n\to0}D_c^{\mathrm{Bethe}}(z=4,n)=1.
\end{equation}
Thus the Bethe-level estimate based only on the local tree instability gives a trivial critical bond dimension in the RTN replica limit.
This result should not be interpreted as the nontrivial finite-dimensional critical point of the square-lattice RTN model, for which a scale around $D_c\simeq2$ has been discussed from the duality viewpoint~\cite{Ohzeki2024}.
Rather, it shows a limitation of the present tree approximation after analytic continuation to $n\to0$.
This observation should be read as a preliminary warning about improving the bulk calculation itself.
Even if one replaces the Bethe approximation by a local cluster approximation such as a plaquette or Kikuchi approximation, the calculation still probes the instability of a homogeneous bulk message unless boundary-condition sectors are introduced explicitly.
Therefore Cayley-tree or Bethe-type estimates of ordinary bulk criticality, although useful for finite-$n$ permutation ferromagnets, should be used with care when they are compared with the RTN entanglement transition.
In the RTN problem, however, the natural object is not the spontaneous bulk magnetization but the free-energy cost of a twist imposed at the boundary.
In other words, one should compare
\begin{equation}
  \Delta F_{\mathrm{twist}}(D)
  =
  F_{\mathrm{twist}}(D)-F_{\mathrm{untwist}}(D),
  \label{eq:twist_free_energy_difference}
\end{equation}
or the associated domain-wall tension, rather than only the bulk order parameter.
This is close in spirit to Peierls-contour or domain-wall free-energy estimates used for random-bond Ising models and surface-code thresholds~\cite{Dennis2002,Honecker2001,Kawashima1997,Amoruso2003}.
For example, in random-bond Ising models, domain-wall free energies under changed boundary conditions have been used to locate the Nishimori point and to diagnose the zero-temperature stability of ordered phases.
In surface-code threshold problems, the relevant comparison is likewise between different error or homology sectors rather than a uniform bulk magnetization.
These examples suggest that, for RTNs, a calculation that keeps track of boundary twists may capture the entanglement transition more faithfully than a calculation based only on the homogeneous bulk cavity fixed point.

At the crudest contour level, a domain wall $\Gamma$ forced by a twist boundary condition has a weight of the form
\begin{equation}
  w(\Gamma)\sim D^{-\Delta_\tau |\Gamma|},
  \qquad
  \Delta_\tau=C(\id)-C(\tau),
  \label{eq:domain_wall_weight}
\end{equation}
where $\tau$ is the relative permutation across the wall.
If the number of contours of length $\ell$ grows as
\begin{equation}
  N_\ell\sim \mu^\ell,
  \label{eq:contour_entropy}
\end{equation}
then the corresponding Peierls-type tension is estimated as
\begin{equation}
  \Sigma(D)
  \simeq
  \Delta_\tau\log D-\log\mu.
  \label{eq:peierls_tension}
\end{equation}
After normalizing by the twist charge $\Delta_\tau$, the directed-path estimate $\mu=2$ gives the rough criterion $D_c\simeq2$.
This domain-wall estimate is a different approximation from the bulk Bethe instability discussed above.
Loop effects, plaquette/Kikuchi counting, or a more refined finite-dimensional renormalization procedure \cite{Parisi2006, Montanari2005, Ohzeki2013bethe} are required to make this boundary-sector estimate quantitative in the RTN limit.

\section{Numerical verification}
We check the analytic criteria by enumerating all elements of $\Sn$ and iterating Eq.~\eqref{eq:cavity} directly.
The initial perturbation is chosen along the character of the standard representation,
\begin{equation}
  \chi^{[n-1,1]}(g)=c_1(g)-1.
\end{equation}
Table~\ref{tab:numerical_check} shows the instability points checked by direct cavity iteration with $z=3$, perturbation amplitude $10^{-6}$, and 80 iterations.
For the cycle-count model, the plotted inverse-temperature variable is $\beta=\log D$.
\begin{table}[tbp]
  \centering
  \small
  \renewcommand{\arraystretch}{1.25}
  \caption{Analytic instability points checked by cavity iteration for $z=3$.}
  \label{tab:numerical_check}
  \begin{tabular}{lrr}
  \toprule
  Model & $n$ & $\beta_c$  \\
  \midrule
  Ising model on $S_2$ & $2$ & $0.549306144334$  \\
  $D^{c(g)}$ & $4$ & $1.609437912434$  \\
  $\exp(\beta c_1(g))$ & $4$ & $0.980748278752$  \\
  $\exp(\beta\delta_{g,\id})$ & $4$ & $3.218875824868$  \\
  \bottomrule
  \end{tabular}
\end{table}

We also monitor two order parameters.
The first is the Potts-like magnetization based on concentration at the identity:
\begin{equation}
  m_{\id}(\eta)
  =
  \frac{n!\eta(\id)-1}{n!-1}.
\end{equation}
The second is the standard-representation magnetization:
\begin{equation}
  m_{\mathrm{std}}(\eta)
  =
  \frac{1}{n-1}
  \sum_{g\in\Sn}\eta(g)(c_1(g)-1).
\end{equation}
Both vanish for the uniform distribution and equal one for a delta mass at the identity.

\begin{figure}[tbp]
  \centering
  \includegraphics[width=0.86\linewidth]{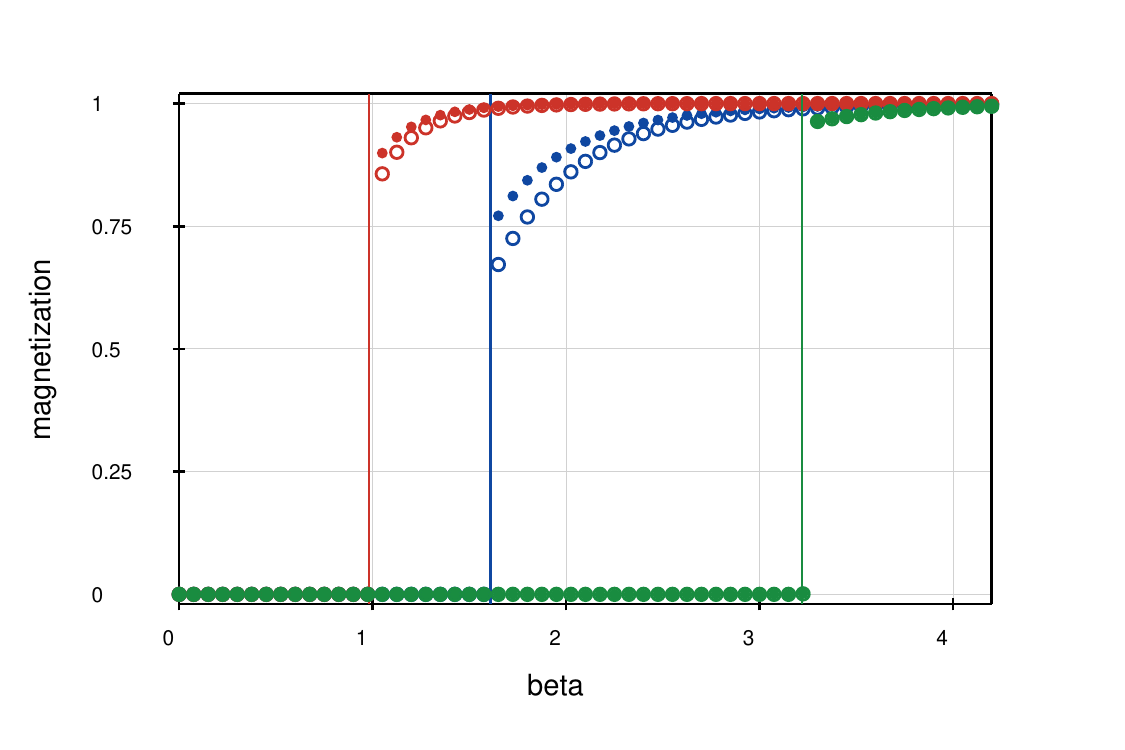}
  \caption{(Color online) Magnetization of the cavity fixed point for $n=4$ and $z=3$. Blue, red, and green points correspond to the cycle-count model $D^{C(g)}$, the fixed-point-enhanced model $\exp(\beta c_1(g))$, and the complete-matching model $\exp(\beta\delta_{g,\id})$, respectively. For each color, filled circles show $m_{\mathrm{std}}$, while open circles show $m_{\id}$. The vertical line of the same color marks the analytic instability point $\beta_c$. In the complete-matching model the filled and open green circles overlap almost completely in the magnetized phase.}
  \label{fig:magnetization-n4k2}
\end{figure}

Figure~\ref{fig:magnetization-n4k2} shows the magnetization of the cavity fixed point as a function of $\beta$.
The fixed-point values are plotted as discrete data points and are not connected by interpolation lines.
The vertical lines indicate the analytic instability points obtained from the standard representation.
Below the instability point the magnetization flows to zero, whereas above it the iteration converges to a magnetized fixed point.
For the complete-matching model, the two plotted order parameters almost coincide in the magnetized phase because the fixed point is essentially Potts-like: one permutation is selected and the remaining permutations are nearly equivalent.
It should be stressed that this is a deterministic iteration of the homogeneous cavity equation from a small standard-representation perturbation.
Thus the numerical check verifies the bulk instability of the finite-$n$ permutation spin model.

\section{Conclusion}
We formulated the cavity method for permutation models on a Cayley tree and analyzed the instability of the uniform paramagnetic solution.
The main observation is that, when the edge Boltzmann weight is a class function on the symmetric group, the edge convolution operator is diagonalized by the noncommutative Fourier transform.
The linearized cavity equation decomposes into irreducible representation sectors of $\Sn$, and the instability condition is determined by the ratios $\omega_\lambda/\omega_{[n]}$.
For cycle-factorized interactions, the required eigenvalues are Schur-function specializations.
This gives a practical way to compute the Bethe or Cayley-tree transition point without diagonalizing an $n!\times n!$ transfer matrix.

Direct numerical iteration of the cavity equation confirms the analytic instability points for the examples considered here.
When this result is used as a Bethe approximation to a square lattice, one sets $z=4$.
At finite $n$ this gives a simple estimate, but the RTN replica limit $n\to0$ yields $D_c=1$.
Thus the present local tree instability gives only a trivial critical bond dimension in the RTN limit.
The reason is conceptual rather than merely numerical: the calculation diagnoses spontaneous bulk magnetization of a finite-$n$ permutation spin model, whereas the RTN entanglement transition is more naturally characterized by the free-energy cost of changing boundary sectors.
The relevant observable is therefore expected to be a twist-induced domain-wall tension, or equivalently a difference between twisted and untwisted free energies, rather than the bulk magnetization alone.

The calculation demonstrates that noncommutative Fourier analysis and symmetric-group representation theory provide a concrete computational framework for permutation models motivated by RTNs.
This framework can serve as a starting point for more elaborate mean-field, Bethe, or real-space renormalization calculations beyond the Cayley-tree geometry.
For the RTN replica limit, however, a promising next step is to combine the representation-theoretic treatment developed here with Peierls-type contour or domain-wall free-energy calculations under twisted boundary conditions.
Such an extension would put the calculation closer to the observables used in RTN and random-circuit entanglement transitions, where domain walls, rather than homogeneous bulk magnetization alone, determine the physically relevant transition.

\section*{Acknowledgements}

We received financial support from the Cross-ministerial Strategic Innovation Promotion Program (SIP) of the Cabinet Office (No. 23836436).

\appendix

\section{Finite-group Fourier transform and Schur's lemma}
\label{app:finite_group_fourier}

This appendix summarizes the finite-group Fourier transform and explains why convolution by a class function becomes scalar in each irreducible representation sector.
We state the argument for a general finite group $G$ and then specialize to $G=\Sn$.

\subsection{Orthogonality of matrix elements}

Let $\rho^\lambda$ be an irreducible unitary representation of $G$ of dimension $d_\lambda$.
The matrix elements satisfy
\begin{equation}
  \frac{1}{|G|}
  \sum_{g\in G}
  \rho^\lambda_{ij}(g)^*
  \rho^\mu_{ab}(g)
  =
  \frac{1}{d_\lambda}
  \delta_{\lambda\mu}\delta_{ia}\delta_{jb}.
\end{equation}
This orthogonality implies that functions on $G$ can be expanded in irreducible matrix elements.
With the convention
\begin{equation}
  \widehat f(\lambda)=\sum_{g\in G}f(g)\rho^\lambda(g),
\end{equation}
the inversion formula is
\begin{equation}
  f(g)=
  \frac{1}{|G|}
  \sum_{\lambda\in\Irr(G)}
  d_\lambda
  \operatorname{Tr}
  \left[
    \widehat f(\lambda)\rho^\lambda(g^{-1})
  \right].
\end{equation}

\subsection{Central elements and Schur's lemma}

Let $W$ be a class function on $G$:
\begin{equation}
  W(h^{-1}gh)=W(g).
\end{equation}
Define
\begin{equation}
  A_\lambda=\sum_{g\in G}W(g)\rho^\lambda(g).
\end{equation}
For any $h\in G$,
\begin{equation}
  \rho^\lambda(h)^{-1}A_\lambda\rho^\lambda(h)
  =
  \sum_{g\in G}W(g)\rho^\lambda(h^{-1}gh).
\end{equation}
Changing variables to $g'=h^{-1}gh$, we get
\begin{equation}
  \rho^\lambda(h)^{-1}A_\lambda\rho^\lambda(h)
  =
  \sum_{g'\in G}W(hg'h^{-1})\rho^\lambda(g')
  =
  \sum_{g'\in G}W(g')\rho^\lambda(g')
  =
  A_\lambda.
\end{equation}
Therefore $A_\lambda$ commutes with every $\rho^\lambda(h)$.
By Schur's lemma, since $\rho^\lambda$ is irreducible,
\begin{equation}
  A_\lambda=\omega_\lambda I_{d_\lambda}.
\end{equation}
Taking the trace gives
\begin{equation}
  \omega_\lambda d_\lambda
  =
  \operatorname{Tr}A_\lambda
  =
  \sum_{g\in G}W(g)\chi^\lambda(g),
\end{equation}
and hence
\begin{equation}
  \omega_\lambda
  =
  \frac{1}{d_\lambda}
  \sum_{g\in G}W(g)\chi^\lambda(g).
\end{equation}

\section{Conjugacy classes of the symmetric group}
\label{app:symmetric_group_classes}

Conjugacy classes of $\Sn$ are classified by cycle type.
Let
\begin{equation}
  \mu=(1^{m_1}2^{m_2}\cdots)
\end{equation}
denote the cycle type with $m_k$ cycles of length $k$.
The centralizer of such a permutation has order
\begin{equation}
  z_\mu=\prod_{k\ge 1}k^{m_k}m_k!.
\end{equation}
Indeed, each cycle of length $k$ admits $k$ cyclic rotations, and the $m_k$ cycles of the same length can be permuted among themselves in $m_k!$ ways.
Multiplying over all $k$ gives $z_\mu$.
By the orbit-stabilizer theorem,
\begin{equation}
  |C_\mu|=\frac{n!}{z_\mu}.
\end{equation}

The permutation representation of $\Sn$ on $\mathbb{C}^n$ has character equal to the number of fixed points:
\begin{equation}
  \chi_{\mathrm{perm}}(g)=c_1(g).
\end{equation}
The one-dimensional subspace spanned by $(1,\ldots,1)$ is the trivial representation.
Its orthogonal complement,
\begin{equation}
  \{(v_1,\ldots,v_n):v_1+\cdots+v_n=0\},
\end{equation}
is the standard representation $[n-1,1]$.
Therefore
\begin{equation}
  \chi^{[n-1,1]}(g)=c_1(g)-1.
\end{equation}

\section{Linearization of the cavity equation}
\label{app:cavity_linearization}

Let
\begin{equation}
  \eta_t=\eta_0+\delta_t,
  \qquad
  \eta_0(\pi)=\frac{1}{|G|},
  \qquad
  \sum_{\pi}\delta_t(\pi)=0.
\end{equation}
Assume that $\delta_t$ lies in a nontrivial irreducible sector $\lambda$.
Then
\begin{equation}
  T_W\eta_t=T_W\eta_0+T_W\delta_t.
\end{equation}
For the uniform part,
\begin{equation}
  T_W\eta_0(\pi)
  =
  \frac{1}{|G|}\sum_{\sigma\in G}W(\pi^{-1}\sigma)
  =
  \frac{\omega_{[n]}}{|G|}
  \equiv a,
\end{equation}
which is independent of $\pi$.
For the nontrivial component,
\begin{equation}
  T_W\delta_t=\omega_\lambda\delta_t.
\end{equation}
Therefore
\begin{equation}
  (T_W\eta_t(\pi))^{z-1}
  =
  (a+\omega_\lambda\delta_t(\pi))^{z-1}
  =
  a^{z-1}+(z-1)a^{z-2}\omega_\lambda\delta_t(\pi)+O(\delta_t^2).
\end{equation}
The denominator becomes
\begin{equation}
  \sum_{\pi}(T_W\eta_t(\pi))^{z-1}
  =
  |G|a^{z-1}
  +
  (z-1)a^{z-2}\omega_\lambda\sum_\pi\delta_t(\pi)
  +O(\delta_t^2)
  =
  |G|a^{z-1}+O(\delta_t^2).
\end{equation}
After normalization, the first-order perturbation is
\begin{equation}
  \delta_{t+1}
  =
  (z-1)\frac{\omega_\lambda}{\omega_{[n]}}\delta_t
  +O(\delta_t^2).
\end{equation}

\section{Derivation of critical condition}
\subsection{Fixed-point-enhanced model}
\label{app:fixed_point_calc}
For the fixed-point-enhanced model,
the generating function of complete homogeneous symmetric functions is
\begin{equation}
  \sum_{m=0}^{\infty}h_m(x) z^m
  =
  \exp\left(\sum_{k=1}^{\infty}\frac{x_k}{k}z^k\right).
\end{equation}
Under this specialization,
\begin{equation}
  \sum_{k=1}^{\infty}\frac{x_k}{k}z^k
  =
  xz+\sum_{k=2}^{\infty}\frac{z^k}{k}
  =
  (x-1)z-\log(1-z).
\end{equation}
Thus
\begin{equation}
  \sum_{m=0}^{\infty}h_m(x) z^m
  =
  \frac{\e^{(x-1)z}}{1-z}.
\end{equation}
Expanding the right-hand side gives
\begin{equation}
  h_m(x)=\sum_{a=0}^{m}\frac{(x-1)^a}{a!}.
\end{equation}

The Jacobi--Trudi formula states
\begin{equation}
  s_\lambda(x)=\det(h_{\lambda_i-i+j}(x))_{1\le i,j\le \ell}.
\end{equation}
For $\lambda=[n-1,1]$,
\begin{equation}
  s_{[n-1,1]}(x)
  =
  \det
  \begin{pmatrix}
    h_{n-1}(x) & h_n(x)\\
    h_0(x) & h_1(x)
  \end{pmatrix}
  =
  h_{n-1}(x)h_1(x)-h_n(x).
\end{equation}
Since $h_0(x)=1$ and $h_1(x)=x$,
\begin{equation}
  s_{[n-1,1]}(x)=xh_{n-1}(x)-h_n(x).
\end{equation}
Using
\begin{equation}
  h_n(x)=h_{n-1}(x)+\frac{(x-1)^n}{n!},
\end{equation}
we find
\begin{equation}
  xh_{n-1}(x)-h_n(x)
  =
  (x-1)h_{n-1}(x)-\frac{(x-1)^n}{n!}.
\end{equation}
With $d_{[n-1,1]}=n-1$, Eq.~\eqref{eq:schur_eigenvalue} gives
\begin{equation}
  \omega_{[n-1,1]}
  =
  \frac{n!}{n-1}
  \left(
    (x-1)h_{n-1}(x)-\frac{(x-1)^n}{n!}
  \right),
\end{equation}
while
\begin{equation}
  \omega_{[n]}=n!h_n(x).
\end{equation}
This proves the expression used in Eq.~\eqref{eq:fixed_point_critical}.

\subsection{Cycle-count model}
\label{app:cycle_count_calc}
For the cycle-count model,
\begin{equation}
  x_k=D\qquad(k\ge 1).
\end{equation}
This is equivalent to evaluating Schur functions at $D$ variables all equal to one, initially for positive integer $D$ and then by polynomial continuation in $D$.
The hook-content formula is
\begin{equation}
  s_\lambda(1^D)
  =
  \prod_{u\in\lambda}
  \frac{D+\cont(u)}{h(u)}.
\end{equation}
The hook-length formula is
\begin{equation}
  d_\lambda
  =
  \frac{n!}{\prod_{u\in\lambda}h(u)}.
\end{equation}
Therefore
\begin{equation}
  \omega_\lambda
  =
  \frac{n!}{d_\lambda}
  s_\lambda(1^D)
  =
  \left(\prod_{u\in\lambda}h(u)\right)
  \left(\prod_{u\in\lambda}\frac{D+\cont(u)}{h(u)}\right)
  =
  \prod_{u\in\lambda}(D+\cont(u)).
\end{equation}

For $[n]$, the contents are
\begin{equation}
  0,1,\ldots,n-1,
\end{equation}
and hence
\begin{equation}
  \omega_{[n]}=D(D+1)\cdots(D+n-1).
\end{equation}
For $[n-1,1]$, the contents are
\begin{equation}
  0,1,\ldots,n-2,-1.
\end{equation}
Thus
\begin{equation}
  \omega_{[n-1,1]}
  =
  (D-1)D(D+1)\cdots(D+n-2).
\end{equation}
Taking the ratio gives
\begin{equation}
  \frac{\omega_{[n-1,1]}}{\omega_{[n]}}
  =
  \frac{D-1}{D+n-1}.
\end{equation}

\bibliographystyle{ptephy}
\bibliography{main}

\end{document}